\documentclass{article}
\usepackage{amsmath,amssymb}
\usepackage{graphicx}
\usepackage{epstopdf}

\numberwithin{equation}{section}

\begin{document}
\title{Buchdahl's inequality in five dimensional Gauss-Bonnet gravity}

\author{
Matthew Wright\footnote{matthew.wright.13@ucl.ac.uk}\\
Department of Mathematics, University College London\\
Gower Street, London, WC1E 6BT, UK
}

\date{\today}

\maketitle

\begin{abstract}
The Buchdahl limit for static spherically symmetric isotropic stars is generalised to the case of five dimensional Gauss-Bonnet gravity. Our result depends on the sign of the Gauss-Bonnet coupling constant $\alpha$. When $\alpha>0$, we find, unlike in general relativity, that the bound is dependent on the stellar structure, in particular the central energy density and we find that stable stellar structures can exist arbitrarily close to the black hole horizon. Thus stable stars can exist with extra mass in this theory compared to five dimensional general relativity. For $\alpha<0$ it is found that the Buchdahl bound is more restrictive than the general relativistic case.
\end{abstract}

\maketitle

\section{Introduction}
An important question in general relativity is determining bounds on the mass and radius of stellar structures. The famous Buchdahl theorem~\cite{Buchdahl:1959zz} says that if we have a static perfect fluid solution to Einstein's equation, whose energy density in non-increasing outwards, then the bound
\begin{align}
\frac{2M}{R}\leq \frac {8}{9}, \label{Buchdahl1}
\end{align}
holds, where $M$ is the mass of the fluid and $R$ is its radius in Schwarzschild coordinates; defined by the location of the vanishing pressure surface. This has a number of important implications. For example, it tells us that the surface redshift is bounded, and that the boundary of a star always occurs after the apparent horizon in the Schwarzschild metric at $r=2M$.

Buchdahl's theorem has been extended in various ways. It has been generalised to include both charge and a cosmological constant, see for example~\cite{Mak:2001gg,HaMa00,Andreasson:2012dj} and references within. Bounds have also been considered without assuming Buchdahl's assumption on non-increasing energy density~\cite{Andreasson:2007ck,Karageorgis:2007cy}.  Such inequalities have also been studied in modified theories of gravity, recently in~\cite{Goswami:2015dma} it was generalised to the case of $f(R)$ gravity. Bounds have also been considered in the context of Brane world scenarios~\cite{Germani:2001du,Garcia-Aspeitia:2014pna}. Studying mass radius bounds in modified theories of gravity is a way of testing their validity. If the theory predicts a bound which is too small or large one can experimentally measure whether there are stars violating these bounds. 

Bounds on the mass radius ratio in higher dimensional general relativity have also been considered by various authors. In $d$-dimensions the natural ratio to consider is
\begin{align}
\frac{2M}{R^{d-3}}
\end{align} 
since this is the component appearing in the higher dimensional Schwarzschild metric, and thus allows one to derive bounds on quantities such as the stars gravitational redshift~\cite{Wright:2015dma}. In~\cite{PoncedeLeon:2000pj} Buchdahl's theorem was extended to $d$-dimensions, where $d\geq4$ and this was generalised to include a non-zero cosmological constant in~\cite{Zarro:2009gd}. Bounds were also considered without assuming Buchdahl's asuumptions in $d$-dimensional spacetimes in~\cite{Wright:2015dma}. 

Gravity in higher dimensions can be extended further than simply considering higher dimensional general relativity. Gauss-Bonnet gravity is a particular natural theory to consider, and appears in the low energy effective action of string theory. This theory is a generalisation of Einstein gravity that adds an extra term to the standard Einstein-Hilbert action, which is quadratic in the Riemann tensor. When varying this extra term with respect to the metric only second order derivatives remain in the field equations, with the higher derivative terms cancelling out exactly, and thus the theory shares many of the nice properties of general relativity. In four dimensions Gauss-Bonnet gravity and general relativity are equivalent, since the Gauss-Bonnet term in the action reduces to a total dervative, giving a surface integral and thus does not add a contribution to Einstein's equation. But when analysing gravity in higher dimensions this extra term is non-trivial and it is thus natural to consider this extra Gauss-Bonnet contribution when considering higher dimensional theories. 

Many authors have considered fluid solutions in the context of Gauss-Bonnet gravity.  Constant density interior solutions were investigated in~\cite{Dadhich:2010qh,Zhou:2011wa}. In~\cite{Hartmann:2013tca} Boson star solutions were considered. Spherical symmetric gravitational collapse has been considered by many authors, see for example\cite{Maeda:2006pm,Jhingan:2010zz} and references therein. 

It has recently been claimed that Buchdahl's theorem is not valid in Gauss-Bonnet gravity~\cite{Hansraj:2015tka}. In this paper we investigate Buchdahl's inequality in five dimensional Gauss-Bonnet gravity. Without the Gauss-Bonnet term the five dimensional Buchdahl inequality derived in~\cite{PoncedeLeon:2000pj} reads
\begin{align}
\frac{2M}{R^{2}}\leq\frac{3}{4}. \label{Buchdahl}
\end{align}
It is of interest to investigate whether this Buchdahl bound, which uniformly bounds the mass radius ratio away from the black hole bound, is a generic result of gravitational theories, or is it in some sense a special property of general relativity. In this paper we use Buchdahl's method~\cite{Buchdahl:1959zz} to show that a version of Buchdahl's inequality does hold in Gauss-Bonnet gravity. In general relativity the interior solution saturating the Buchdahl bound is given by the constant density solution. This is not always the case for the inequality we have derived for the Gauss-Bonnet case. However it is found that for a stable stellar configuration, as in general relativity, the apparent horizon lies strictly below the radius of the star, however one cannot bound this uniformly away from the horizon. 

This paper is organised as follows. In Section II we review Gauss-Bonnet gravity and the five dimensional black hole solution and we explore the consequences on the mass radius bound for the constant density solution. In section III we derive the field equations for a spherically symmetric static perfect fluid and rewrite them in Buchdahl variables. In section IV we make the standard Buchdahl assumptions and derive bounds on the exterior vacuum metric. Finally in section V we use these bounds to derive the corresponding Gauss-Bonnet mass radius bound.

\section{Gauss-Bonnet gravity}

In this section we introduce the action and field equations of Gauss-Bonnet gravity. The action for five dimensional Gauss-Bonnet gravity is given as  follows
\begin{align}
S=\int d^5 x \sqrt{-g} \left(\frac{1}{2\kappa}\left[ R+\alpha L_{\rm GB} \right]\right) +S_{\rm matter}. 
\end{align}
The first term is the usual Einstein Hilbert action, whereas the last is the usual matter Lagrangian. The Gauss-Bonnet Lagrangian is given by the following particular combination of Ricci scalar, Ricci tensor and Riemann tensor
\begin{align}
L_{\rm GB}=R^2-4R^{AB}R_{AB}+R_{ABCD}R^{ABCD}. \label{GB}
\end{align}
This particular Lagrangian appears from the low energy limit of heterotic superstring theory~\cite{Gross:1998jx}. Note that we will not consider the effect of a cosmological constant in this paper. 

Varying the action with respect to the metric gives the following generalisation of Einstein's field equations
\begin{align}
G_{AB}+\alpha H_{AB}=\kappa T_{AB}. \label{Einstein}
\end{align}
Here $G_{AB}$ is the Einstein tensor, $T_{AB}$ is the usual energy momentum tensor, and the tensor $H_{AB}$ is given by
\begin{align}
H_{AB}=2 \left[ RR_{AB}-2R_{AC}R^C_B-2R^{CD}R_{ABCD}+R_A^{CDE}R_{BCDE}\right] -\frac{1}{2} g_{AB}L_{\rm GB}.
\end{align}
From the interpretation of the low energy effective action from string theory, the coupling constant $\alpha$ is related to the inverse string tension which is positive definite, and thus the condition $\alpha\geq 0$ is usually considered. This is also required for stability of Minkowski space in this theory. Nonetheless, in this paper we will also consider the case of $\alpha<0$, which has been considered by some authors, see for example~\cite{Guo:2009uk,Hansraj:2015tka}. We will also assume geometric units $\kappa=1$ from now on.

\subsection{Gauss-Bonnet Black hole}

Assuming a static spherically symmetric metric and solving the field equations gives the following five dimensional black hole solution, first derived in~\cite{Boulware:1985wk}
\begin{align}
ds^2=-F(r)dt^2+\frac{dr^2}{F(r)}+r^2\left(d\theta^2+\sin^2 \theta \, (d\varphi^2+\sin^2 \varphi  \, d\psi^2)\right) \label{GBblackhole},
\end{align}
where the function $F(r)$ is given by
\begin{align}
F(r)=1+\frac{r^2}{4\alpha}\left(1-\sqrt{1+\frac{16 \alpha M}{r^4}}\right)
\end{align}
where $M$ is related to the mass of the black hole or interior. Taking the limit $\alpha \rightarrow 0$ in this expression we recover the five dimensional Schwarzschild metric
\begin{align}
F(r)=1-\frac{2M}{r^2}.
\end{align}

For this solution to describe the exterior of a star we require the apparent horizon to exist before the boundary of the star at $r=R$. The event horizon of a Gauss-Bonnet black hole is located at
\begin{align}
R=\sqrt{2M-2\alpha}
\end{align} 
This gives the following condition on the mass radius ratio
\begin{align}
\frac{2M}{R^2} \leq 1+ \frac{2\alpha}{R^2}, \label{EH}
\end{align}
assuming $\alpha>-R^2/4$. If $\alpha<-R^2/4$ then we simply have the condition
\begin{align}
\frac{2M}{R^2}\leq \frac{1}{2}
\end{align}
For positive $\alpha$ this mass-radius bound is less strict than the pure general relativistic case.

Birkhoff's theorem in five dimensional Gauss-Bonnet gravity does not hold due to the presence of branch cuts, and thus this exterior solution is not unique. However we will only consider interior solutions which match to this exterior solution in this paper, as this is the branch which is asymptotically flat and agrees with the Schwarzschild metric in the limit $\alpha\rightarrow 0$.

\subsection{Constant density solution}

In standard general relativity the constant density solution saturates the Buchdahl bound in any dimension. The constant density interior solution for five dimensional Gauss-Bonnet gravity was derived in~\cite{Dadhich:2010qh}. We review this solution here and discuss the implications on the mass radius ratio.  

The energy density of the solution was taken to be a constant $\rho$ and the pressure in the interior is given by
\begin{align}
p=\frac{3}{4\alpha}(1-\mu)\left[1-\frac{\mu}{1+\frac{2A \sqrt{\alpha}}{B\sqrt{r^2(1-\mu)+4\alpha}}}\right] \label{pressure}
\end{align} 
where the quantity $\mu$ is defined by
\begin{align}
\mu=\sqrt{1+16\alpha w_b}, \quad w_b= \frac{\sqrt{1+\frac{4}{3} \alpha \rho}-1}{8\alpha}.
\end{align}
The constants $A$ and $B$ and the mass of the fluid are determined by the matching conditions; requiring that the metric components $g_{tt}$ and $g_{rr}$ are continuous at $r=R$, as well ass $g_{tt,r}$ being continuous on this surface. This implies that the mass is given by
\begin{align}
M=\frac{1}{12}\rho R^4
\end{align}
and $A$ and $B$ are given by
\begin{align}
A&=(1-B)\sqrt{F(R)}, \\
B&=-(1+\frac{16\alpha M}{R^4})^{-1/2}.
\end{align}

In general relativity finiteness of central pressure of the constant density solution gives the Buchdahl bound~(\ref{Buchdahl}). Let us examine what this condition gives in the Gauss-Bonnet case. Inserting $r=0$ into~(\ref{pressure}) and requiring this to be finite and positive gives the inequality
\begin{align}
\frac{A}{B}\leq -1. \label{AB}
\end{align}
This allows us to derive the following condition on $F$
\begin{align}
F(R) \geq \frac{1}{4(1+4\alpha w_b)^2}. \label{Fboundc}
\end{align}
The right hand side of this is well defined even in the case of $\alpha<0$, since $1+4\alpha w_b>0$. We will discuss the implications of this bound on the mass radius ratio in Section~\ref{sec6}.

\section{Field Equations for static star}

In this section we will consider the solutions for a more general perfect fluid. We assume a spherically symmetric static metric of the form
\begin{align}
ds^2=-e^\nu dt^2+e^\lambda dr^2+r^2 d\Omega_3^2
\end{align}
where $d\Omega_3^2$ is the metric of a $3$-sphere and the functions $\mu$ and $\lambda$ depend only on the radial coordinate $r$. We take the energy momentum tensor to be that of a perfect fluid
\begin{align}
T^A_B = {\rm diag}(-\rho,p,p,p,p)
\end{align}
where $\rho$ is the energy density of the fluid and $p$ is the isotropic pressure. The additional component of  the energy momentum tensor in five dimensions, $T^4_4$ is required to be identical to $T^2_2$ and $T^3_3$ under the assumption of spherical symmetry. Working in natural units with $\kappa=1$, the $(t,t)$ component of the field equations~(\ref{Einstein}) gives
\begin{align}
\rho= \frac{3e^{-\lambda}}{2r^2} \left( r \lambda'-2(1-e^\lambda)\right)-\frac{6\alpha\lambda' e^{-2\lambda }}{r^3}(1-e^\lambda) \label{gtt}
\end{align}
while the $(r,r)$ component reads
\begin{align}
p=\frac{3 e^{-\lambda}}{2r^2}\left(r \nu' +2(1-e^\lambda)\right)-\frac{12\alpha \nu' e^{-\lambda}}{r^3}(1-e^\lambda). \label{grr}
\end{align}
Conservation of the energy momentum tensor will give the following equation
\begin{align}
p'=-(\rho+p)\nu' \label{conservation}
\end{align}
which is unmodified from general relativity. If these three equations are satisfied, the remaining Einstein equations, that is the $\theta-\theta$, $\phi-\phi$ and $\varphi-\varphi$ components, are identically satisfied, as in the standard four dimensional case.

Let us write $e^{-\lambda}=1-f(r)$. From equation~(\ref{gtt}) we get
\begin{align}
(2\alpha f^2 +r^2 f)'=\frac{2}{3} \rho(r) r^3 \label{density}
\end{align}
and hence integrating this gives
\begin{align}
2\alpha f^2 +r^2 f=\frac{2}{3}\int \rho(r') r'^3 dr'.
\end{align}
Let us introduce the mass function
\begin{align}
 m=\frac{1}{3}\int_{0}^{r} \rho(r') r'^3 dr',
\end{align}
where the factor of $1/3$ at the front required as the higher dimension mass function possesses an additional factor of $1/(d-2)$~\cite{PoncedeLeon:2000pj}. By matching this interior solution with the exterior vacuum metric~(\ref{GBblackhole}) we see that evaluating this mass function at the boundary of the star gives the total mass of the fluid
\begin{align}
M=m(R).
\end{align}
Now we can solve for $f$ to find the metric function $e^{-\lambda}$
\begin{align}
e^{-\lambda}=1-f=1-r^2\left(\frac{\sqrt{1+\frac{16\alpha m(r) }{r^4}}-1}{4\alpha}\right)
\end{align}

Now let us define our Buchdahl variables. First we introduce the function $w(r)$ such that  
\begin{align}
e^{-\lambda}=1-2r^2 w(r)
\end{align}
which we can solve for $w$ to give
\begin{align}
w(r)=\frac{\sqrt{1+\frac{16 \alpha m(r)}{r^4}}-1}{8\alpha}. \label{w}
\end{align}
We also introduce the following further variables $x$, $y$ and $\zeta$:
\begin{align}
x=r^2, \quad \zeta= e^{\nu/2}, \quad y^2=e^{-\lambda}=1-f(r).
\end{align}

In terms of these new variables we can then rewrite the $(r,r)$ component of Einstein's equation~(\ref{grr}) as follows 
\begin{align}
p=6y^2 \frac{\zeta_{,x}}{\zeta}-6 w+48 \alpha y^2 w\frac{\zeta_{,x}}{\zeta}. \label{pressure1}
\end{align}
Rewriting equation~(\ref{density}) in terms of $w$ instead of $f$ yields the following equation
\begin{align}
\rho=6(x w_{,x}+2w+8\alpha w(x) (x w_{,x}+w)).
\end{align}
Inserting this into the conservation equation~(\ref{conservation}) gives
\begin{align}
p_{,x}=-(6(x w_{,x}+2w+8\alpha w (x w_{,x}+w))+p)\frac{\zeta_{,x}}{\zeta}. \label{conservation2}
\end{align}

By differentiating~(\ref{pressure1}) with respect to $x$ and inserting this into~(\ref{conservation2}) we can eliminate the pressure from these equations, which after simplification gives the following equation
\begin{align}
((1+8\alpha w)y \zeta_{,x})_{,x}-\frac{w_{,x} \zeta}{y}=0. \label{diff1}
\end{align}
Let us now introduce the final Buchdahl variable $\xi$ given implicitly by
\begin{align}
d\xi=\frac{dx}{y}.
\end{align}
We can now use this to rewrite~(\ref{diff1}) as the following
\begin{align}
((1+8 \alpha w)\zeta_{,\xi})_{,\xi}-w_{,x} \xi =0 .\label{zetadiff}
\end{align}

\section{Buchdahl bounds}

We are now in a position to derive our main results. In this section we will derive bounds on the metric function $F$ at the boundary $r=R$. We will assume the standard Buchdahl assumption that the energy density is a monotonically decreasing function with respect to $r$. Thus we see from differentiating~(\ref{w}) that
\begin{align}
w'(r)=\frac{r m'(r)-4m(r)}{r^5 \sqrt{1+\frac{16 \alpha m(r)}{r^4}}}\leq 0 \label{wdecreasing}
\end{align}
In what follows we will denote quantities evaluated at the centre $r=0$ with a subscript $c$ and quantities at the boundary $r=R$ with a subscript $b$.

Now from inequality~(\ref{wdecreasing}) we have
\begin{align}
w_{,x}\leq 0.
\end{align}
Inserting this into~(\ref{zetadiff}) yields the following relation
\begin{align}
((1+8 \alpha w)\zeta_{,\xi})_{,\xi}=w_{,x} \zeta \leq 0 .
\end{align}
Thus we can deduce that
\begin{align}
(1+8 \alpha w)\zeta_{,\xi} \geq (1+8 \alpha w_b)(\zeta_{,\xi})_b. \label{zetaineq1}
\end{align}

Now evaluating $\zeta_{,\xi}$ at the boundary, using that $\zeta$ is $\mathcal{C}^1$ at the boundary so matches the derivative of the exterior solution, yields
\begin{align}
\zeta_{,\xi} \mid_b= \frac{F'(R)}{4R}=\frac{ w_b}{1+8\alpha w_b}.
\end{align}
Inserting this back into~(\ref{zetaineq1}) thus gives 
\begin{align}
(1+8 \alpha w)\zeta_{,\xi} \geq w_b
\end{align}
Integrating both sides of this with respect to $\xi$ gives
\begin{align}
\int_{0}^{\xi_b} (1+8 \alpha w)\zeta_{,\xi}\, d\xi \geq 2w_b \int_{0}^{R} \frac{r}{y}\, dr \label{ineq1}
\end{align}
Now we find a lower bound for the left hand side of~(\ref{ineq1}). Using the fact $w$ is decreasing means $y\leq \sqrt{1-2w_b r^2}$ and hence
\begin{align}
\int_{0}^{R} \frac{r}{y}\, dr &\geq \int_{0}^{R}\frac{r}{\sqrt{1-2 w_b r^2}} dr \\ &= \frac{1}{2w_b}(1-\sqrt{1-2 w_b r^2})
\\& =\frac{1}{2w_b} (1-\sqrt{F(R)}). \label{ineq3}
\end{align}
Now to examine the left hand side of~(\ref{ineq1}) we will need to consider the cases of positive and negative $\alpha$ separately.

\subsection{Bounds for$\alpha>0$}

First we will consider the more physically relevant case $\alpha>0$. Now since $w$ is a decreasing function, (\ref{zetaineq1}) is a weaker inequality than the corresponding one in general relativity; and it will no longer necessarily be saturated by the constant density solution.

Now we wish to find an upper bound for the left hand side of~(\ref{ineq1}). Since $\alpha$ is positive and $w$ is decreasing, $w$ can be bounded by its central value
\begin{align}
\int_{0}^{\xi_b} (1+8 \alpha w)\zeta_{,\xi} d\xi &\leq \int_{0}^{\xi_b} (1+8 \alpha w_c) \zeta_{,\xi} d\xi
\\
&\leq (1+8 \alpha w_c) \zeta_b
\\
&=(1+8 \alpha w_c) \sqrt{F(R)} \label{ineq2}
\end{align}
And hence going back to~(\ref{ineq1}) and using~(\ref{ineq2}) and~(\ref{ineq3}) we find the following inequality holds 
\begin{align}
\sqrt{F(R)}(1+8 \alpha w_c) \geq  (1-\sqrt{F(R)}).
\end{align}
Rearranging this to find $F(R)$ gives the following bound on the metric function
\begin{align}
F(R) \geq \frac{1}{4 (1+4\alpha w_c)^2} .\label{Fbound}
\end{align}
In the case of a constant density solution, we have that $w_c=w_b$, and hence our inequality~(\ref{Fbound}) will in this case agree with the bound~(\ref{Fboundc}) found in the constant density case.

\subsection{Bounds for $\alpha<0$}

Now we consider the case $\alpha<0$. We must impose the condition $1+\frac{16\alpha M}{R^4}>0$ in order for the exterior solution to be well defined, which implies $1+8\alpha w_b>0$. Now, since $w$ is decreasing, this time we can bound the left hand side of~(\ref{ineq1}) using the boundary value of $w$
\begin{align}
\int_{0}^{\xi_b} (1+8 \alpha w)\, \zeta_{,\xi}\, d\xi &\leq (1+8 \alpha w_b)\int_{0}^{\xi_b} \zeta_{,\xi} d\xi \\
& \leq (1+8 \alpha w_b) \sqrt{F(R)}
\end{align}
A lower bound on the right hand side of~(\ref{ineq1}) was already obtained, and thus we find
\begin{align}
(1+8 \alpha w_b) \sqrt{F(R)}\geq  (1-\sqrt{F(R)}).
\end{align}
Hence we can derive the following bound on the metric function $F(R)$
\begin{align}
F(R) \geq \frac{1}{4 (1+4\alpha w_b)^2} .\label{Fbound2}
\end{align}
This agrees with the bound~(\ref{Fboundc}) that was found for the constant density solution. 

\section{Mass-radius ratio bounds} \label{sec6}
In this section we will examine what the implications of the bounds~(\ref{Fbound}),(\ref{Fbound2}) on the metric function $F(R)$ have on the mass radius ratio.

First we will examine the case of $\alpha>0$, where the inequality~(\ref{Fbound}) was derived. Let us define the quantity
\begin{align}
\delta_{\alpha}:= 1+4\alpha w_c \geq 1.
\end{align}
Now we can rewrite $w_c$ in terms of the central energy density as follows
\begin{align}
w_c= \lim\limits_{r\rightarrow 0}\frac{\sqrt{1+\frac{16\alpha m}{r^4}}-1}{8\alpha}=\frac{\sqrt{1+\frac{4\alpha \rho_c}{3}}-1}{8\alpha}
\end{align}
Then rearranging~(\ref{Fbound}) allows us to derive the following bound on the mass-radius ratio
\begin{align}
\frac{2M}{R^2} \leq (1-\frac{1}{4\delta_\alpha^2})+\frac{2\alpha}{R^2}(1-\frac{1}{2\delta_\alpha^2}+\frac{1}{16 \delta_\alpha^4}). \label{GBBuchdahl}
\end{align}
This is strictly weaker than the corresponding five dimensional general relativistic Buchdahl bound. The right hand side of this inequality can in principle approach the horizon bound if the central energy density is large. However, this result is not necessarily saturated by any solution, as we have bounded the left hand side and the right hand side of~(\ref{ineq1}) by different solutions. Thus in order to examine whether the bound does get arbitrarily close to the horizon bound, we will examine the constant density solution again.

In the case of both the constant density solution for both positive and negative $\alpha$, and the generic case of $\alpha<0$, the same bounds~(\ref{Fbound2}),(\ref{Fboundc}) were derived
\begin{align}
F(R) \geq \frac{1}{4 (1+4\alpha w_b)^2} \label{Fboundnc}
\end{align}
Now since the right hand side involves $w_b$, which is given explicitly by 
\begin{align}
w_b= \frac{\sqrt{1+\frac{16 M\alpha}{R^4}}-1}{8\alpha}
\end{align}
finding an inequality on the mass-radius ratio will reduce to solving a cubic equation. Rearranging~(\ref{Fboundnc}) gives the following cubic equation in $w_b$ which must be satisfied
\begin{align}
H(w_b):=-128 \alpha ^2 R^2 w_b^3+64 \alpha \left(\alpha -R^2\right) w_b^2 -8\left(R^2-4 \alpha \right) w_b +3\geq0~\label{cubic1} 
\end{align} 
Let us introduce the variables
\begin{align}
f=2 R^2 w_b, \quad \beta=\frac{\alpha}{R^2}.
\end{align}
In the limit $\alpha \rightarrow 0$, $f=2M/R^2$. Now the inequality~(\ref{cubic1}) becomes
\begin{align}
H(f)=-16 \beta ^2 f^3+16 \beta(\beta-1) f^2+4( 4\beta  -1)f+3\geq0
\end{align}

For positive $\beta$ the cubic has a positive discriminant and hence has three real roots. Only one of these roots is positive. One can see this by noting  that the cubic is negative, $H(-1/2\beta)=-1<0$ and $H(0)=3>0$ and hence two of the roots must be negative. Let us call this positive root $\gamma$. Now since the horizon occurs and $f=1$, and $H(1)=-1<0$ we readily see that we must have $\gamma<1$, and thus the bound on $f$ lies strictly below the horizon.

For negative $\beta$, the discriminant of this cubic is negative when approximately $-1.32343\leq \beta<0$. In this range the cubic has only one real root, which is positive, since the cubic is negative and $H(0)=3>0$. We will denote this root by $\gamma$ also.  For $\beta<-1.32343$ the cubic has three real roots, all of which are positive. However, we note for negative $\beta$ the inequality
\begin{align}
1+2\beta f> 1+4 \beta f=\sqrt{1+\frac{16\alpha M}{R^4}}>0
\end{align}
holds, and thus $f\leq -1/2\beta$. Now, $H(-1/2\beta)=-1<0$,  $H'(-1/2\beta)=0$, and thus the cubic has a turning point at this value. By looking at the second deivative of $H$ we see that this turning point is a minimum for $\beta<-1.32343$ meaning that two of the roots of the cubic are above $-1/2\beta$. Thus only one of the roots of the cubic is less than $-1/2\beta$, and we will also denote this root again by $\gamma$. Hence for all values of $\beta$ the cubic inequality reduces to the simple condition $f \leq \gamma$.

\begin{figure}[!ht]
\centering
\includegraphics[width=0.48\textwidth]{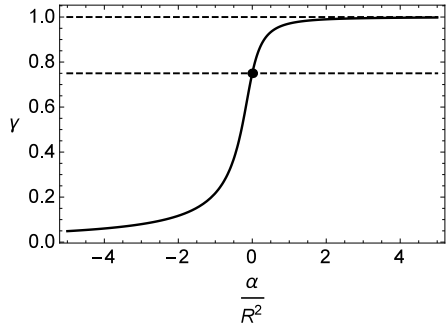}
\caption{Plot of the root $\gamma$ against $\beta:=\alpha/R^2$. The top dashed line indicates the horizon, whereas the lower dashed line indicates the general relativistic bound. The point shows the location of vanishing $\alpha$, agreeing with the five dimensional general relativistic result. Positive $\beta$ gives a less stricter bound than in general relativity, and approaches the horizon as $\beta$ increases, whereas negative $\beta$ gives a stricter bound and approaches $0$ as $\beta$ decreases.}\label{zgraph}
\end{figure}

We plot the graph of this positive root $\gamma$ of this cubic in Fig~\ref{zgraph}. We note that $\gamma$ varies continuously with $\beta$; the root $\gamma$ belongs to the same branch cut. We see that for $\alpha>0$ the result rapidly approaches the horizon bound at $w_b=1$. For negative $\alpha$ the bound on $\gamma$ is much stricter than the general relativistic case.

Now this root of the cubic $\gamma$  does not have a pleasing analytic form. However, in terms of $\gamma$ we can find the following bound on the mass-radius ratio
\begin{align}
\frac{2M}{R^2} \leq \gamma (1+\frac{2\alpha \gamma}{R^2}) \label{gammab}
\end{align}
for $R^2>-4\alpha \gamma$, whereas for  $R^2<-4\alpha \gamma$ we have
\begin{align}
\frac{2M}{R^2} \leq \frac{\gamma}{2} \label{gammab2}
\end{align}
This is less than the horizon bound, but $\gamma\rightarrow 1$ as $\alpha/R^2$ gets large, and therefore there is no uniform bound below the horizon for $\alpha>0$. 

But for $\alpha<0$ we can find some simple results on the mass radius bound. To do this we note a sufficient condition for~(\ref{Fboundnc}) to be true is that
\begin{align}
F(R)\geq \frac{1}{4}.
\end{align}
This is equivalent to noting that in this case $\gamma<3/4$. Analysing this weaker bound alone allows us to draw important conclusions. Rearranging this, we find the two following bounds on the mass radius ratio, dependent on the radius of the star. For $R^2<-3\alpha$ we have
\begin{align}
\frac{2M}{R^2}\leq \frac{3}{8} \label{GBBuchdahl2}
\end{align}
and for $R^2>-3\alpha$ we have
\begin{align}
\frac{2M}{R^2} \leq \frac{3}{4}+ \frac{9}{8R^2}\alpha \label{GBBuchdahl3}
\end{align} 

\section{Discussion}

In this paper we have investigated the equivalent of the Buchdahl bound in five dimensional Gauss-Bonnet gravity. For positive coupling constant $\alpha$ we have derived the inequality~(\ref{GBBuchdahl}).  Unlike in five dimensional general relativity, the bound is not independent of the stellar structure since it depends on the central energy density through the value of $\rho_c$. However, one important consequence of our bound is that it is strictly less than the position of the event horizon, in particular
\begin{align}
\frac{2M}{R^2}< 1+\frac{2\alpha}{R^2}. \label{horizon}
\end{align}
This is because the quantity $\delta_\alpha$ is finite and greater than $1$. This means that Gauss-Bonnet black holes cannot have a perfect fluid interior either.

In principle the right hand side of the inequality~(\ref{GBBuchdahl}) can approach the horizon bound~(\ref{horizon}) if the central density becomes large. Thus unlike in general relativity we cannot give a generic bound on quantities such as the surface redshift. In five dimensional general relativity, the surface redshift $z$ is bounded above by $z\leq 1$. However our result shows that the redshift bound will be dependent on the specific stellar structure, agreeing with the conclusion in~\cite{Zhou:2011wa}. Also, unlike in general relativity, the constant density solution does not saturate the bound~(\ref{Fbound}). However, even analysing the constant density solution, we see that this can be arbitrarily close to the black hole bound, and so we can see that this black hole bound is saturated by a fluid solution in Gauss-Bonnet gravity.

In the limit $\alpha \rightarrow 0$ of this inequality we recover five dimensional general relativistic Buchdahl bound
\begin{align}
\frac{2M}{R^2}\leq \frac{3}{4}.
\end{align}
The Gauss-Bonnet inequality is less restrictive than this general relativistic bound, and thus we are able to conclude that the appearance of the Gauss-Bonnet term allows stable configurations of stars with more mass in a given radius than their general relativistic counterparts. 

On the other hand, considering negative $\alpha$ gives a completely different situation.  In this case these the constant density solution does saturate the bound~(\ref{Fbound2}) on the function $F(R)$. The weaker inequality~(\ref{Fboundnc}) is also true, and this allows us to derive the explicit bounds~(\ref{GBBuchdahl2}) and (\ref{GBBuchdahl3}), and are valid depending on the size of the radius of the star. Both of these bounds are stricter than the general relativistic bound, and are independent of the exact stellar structure. Again, the bound~(\ref{GBBuchdahl3}) correctly reduces to the standard five dimensional Buchdahl bound in the limit $\alpha \rightarrow 0$. This means we can fit less mass into a given radius and while maintaining a stable stellar structure. These inequalities will also imply a smaller upper bound for the surface redshift. The right hand side of inequality~(\ref{gammab2}) approaches $0$ as the ratio $\alpha/R^2$ decreases, and thus as $\alpha$ or $R$ decreases the mass-radius ratio gets very small. 

Of further interest would be to consider the maximum mass of a neutron star in Gauss-Bonnet gravity. In general relativity, it is known that $M_{\rm max}<3.2 M_{\odot}$~\cite{Rhoades:1974fn}, with $M_{\odot}$ the solar mass, where certain physical assumptions were made on the structure of the neutron star interior. Such an analysis has been done for higher dimensional general relativity for one particular EoS in~\cite{Bordbar:2015wva}, where it was found that in higher dimension the neutron star maximum mass violated the Schwarzschild bound, and thus did not in fact describe a neutron star. Thus a further investigation of the structure of higher dimensional neutron stars in both general relativity and Gauss-Bonnet gravity should take place. 

\section*{Acknowledgements}
The author would like to thank Christian B\"{o}hmer for useful discussions and helpful comments on the manuscript.

\end{document}